\documentclass[aps,prd,amsmath,amssymb,superscriptaddress,showpacs]{revtex4}

\usepackage{epsfig,graphics,color}
\usepackage{graphicx}
\usepackage{dcolumn}
\usepackage{bm}
\begin{document}

\title{Hadronic interactions of the $J/\psi$ and Adler's theorem}

\author{A. Bourque}
\affiliation{Department of Physics, McGill University\\ 
3600 University Street, Montreal, QC, 
Canada H3A 2T8}
\author{C. Gale}
\affiliation{Department of Physics, McGill University\\  
3600 University Street, Montreal, QC, 
Canada H3A 2T8}
\author{K. L. Haglin}
\affiliation{Department of Physics \& Astronomy, Saint Cloud State University,
720 Fourth Avenue South, St. Cloud, MN 56301, USA}

\date{ \today}

\begin{abstract}
Effective Lagrangian models of charmonium have recently been used to
estimate dissociation cross sections with light hadrons.
Detailed study of the symmetry properties reveals possible
shortcomings relative to chiral symmetry.
We therefore propose a new Lagrangian and point out distinguishing 
features amongst the different approaches.  Moreover, we test the models against
Adler's theorem, which requires, in the appropriate limit, the decoupling of 
pions from the theory for the normal parity sector. 
Using the newly proposed Lagrangian, which 
exhibits $SU_L(N_f)\times\,SU_R(N_f)$ symmetry and complies with
Adler's theorem, we find dissociation cross sections with pions that are 
reduced in an energy dependent way, with respect to cases where the theorem is
not fulfilled.
\end{abstract}


\maketitle


\section{Introduction}


The theoretical study of matter under extreme conditions enjoys a wide range of 
application, from the physics of the early Universe, to that of relativistic 
nuclear collisions. The latter offer the tantalizing possibility of recreating 
in the laboratory the conditions that prevailed roughly a microsecond after the
Big Bang. The theory of the strong interaction, Quantum Chromodynamics (QCD),
predicts a phase transition from normal hadronic matter to a plasma of quarks
and gluons \cite{kl03}. To find a signature of this new state of matter
represents a task which has generated tremendous activity both in theory and in
experiment. Of the many signals put forward as probes of the quark-gluon plasma,
the suppression of the $J/\psi$ yield enjoys a popular status \cite{mat86}. 

Indeed,
since charmonium is predominantly produced in the early stage of the nuclear
collisions through hard processes, it acts as a probe for the subsequent stages.
The original idea was that the presence of  a quark-gluon plasma (QGP) will screen the
long-range confining force between $c$-$\bar c$, leading to the decoherence of
the pair \cite{mat86}. This suppression mechanism was later augmented by the
possibility of charmonium dissociation by hard gluons in a deconfined medium \cite{ks94}.
In the interpretation of the early centrality-dependent $J/\psi$ absorption
observed by the NA38 collaboration \cite{na38}, those suppression mechanisms
were not manifest and nuclear absorption sufficed to understand the data. 
But, this scenario
says nothing about the effects of the late hadronic phase. Indeed, also accounting for 
final state interactions can go a long way in reproducing the NA50
suppression pattern \cite{na50} observed subsequently in Pb + Pb collisions, provided 
the $J/\psi$
cross-sections with hadronic matter are of the order of one to a few 
millibarns \cite{arm98,cap02}. In those experiments, it is fair to say that the
presence of a quark-gluon plasma is still ambiguous. 
Thus, in order to identify the true nature of a possibly new phase, it appears necessary to
quantify the $J/\psi$-light hadron interaction.
This requirement on the $J/\psi$ cross-sections with light mesons
is not a trivial one to satisfy, as there are no direct experimental measurements. One
has to rely on theoretical calculations based, for example, on QCD sum-rules
\cite{dur03}, on quark-potential models \cite{mar95,won01}, or on effective
mesonic Lagrangians \cite{mat95, kh00, hag00, lin00, oh01}. These lead
typically to cross-sections from a few tenths of a millibarn to a few
millibarns near threshold \cite{dur03}. 

Specifically, for the
effective mesonic Lagrangians found in \cite{hag00, lin00, oh01}, once form
factors have been folded-in  to account for short-range interactions, the
dissociation cross-section by pions reaches a few millibarns. But concerns have
been raised about using such models: (i) the $SU(4)$ symmetry used to describe
the pseudoscalar and vector meson interactions is questionable as it is  
broken, (ii) the form factors accounting for the finite size of the mesons do
not proceed from the formalism as in other models \cite{dur03, won01}, and
(iii) the $J/\psi + \pi \rightarrow D^* + \bar D$ process does not vanish 
in the soft-pion
limit for non-degenerate vector meson masses as expected from Adler's theorem 
\cite{nav01}. All of those can be addressed, but it is the purpose of
this article to expand on the last point and show that, even in the degenerate
vector mass limit, the pions' decoupling in the soft limit does not necessarily
follow for amplitudes
constructed from the normal parity content of the Lagrangians found in
\cite{hag00, lin00, oh01}. Consequently, we propose an alternative
charmonium Lagrangian implementing the extended $SU_L(4) \times SU_R(4)$ chiral
symmetry for which the theorem holds for a degenerate vector mass spectrum.
\\\indent Our paper is organized as follows: we first present two models
implementing the $SU_L(N_f)\times SU_R(N_f)$ chiral symmetry and one implementing
only the $SU_V(N_f)$ symmetry. Having introduced the pseudoscalar mesons as
Goldstone bosons in all these models, we note that for $SU_L(N_f)\times SU_R(N_f)$
invariant normal parity Lagrangians, the pseudoscalars must decouple from the
theory in the zero-momentum limit. We explicitly show that at the amplitude
level for the process $\rho^0 + \pi^+ \rightarrow \rho^0 + \pi^+$, the
$SU_L(2)\times SU_R(2)$ chirally symmetric models obey the soft-pion theorem, 
while the isospin-only invariant model does not. We then digress on 
electromagnetic
current conservation, and show that it can be implemented in all the models.
Bearing in mind that we wish to study the implementation of pions' decoupling in
effective charmonium Lagrangians, we review the most commonly used effective
Lagrangians in hadronic $J/\psi$ suppression, and show that they are not
invariant under the $SU_L(4)\times SU_R(4)$ symmetry, but rather only under
$SU_V(4)$. We explore the consequences of implementing the extended chiral
symmetry by comparing the cross-section for the $J/\psi + \pi$ absorption
process obtained in our two formulations. Finally, in Appendix \ref{appA}, we
show that the amplitude for the $J/\psi + \pi$ absorption process does vanish in
the degenerate vector meson mass limit for the extended chiral symmetric case,
but not the $SU_V(4)$. We re-derive the result of Ref.~\cite{nav01} 
for the case of a non-degenerate vector mass spectrum. We then 
explicitly check in Appendix \ref{appB} that the Ward identity holds for both $SU_V(4)$ 
and $SU_L(4)\times SU_R(4)$ models. In Appendix \ref{appC}, a discussion about
fixing coupling constants is presented. 


\section{EFFECTIVE MESONIC LAGRANGIANS}



\subsection{$SU_L(N_f)\boldmath{\mbox{$\times$}} SU_R(N_f)$ Lagrangian
with pseudoscalar, vector and axial vector mesons }

 
To build a $SU_L(N_f)\times SU_R(N_f)$  symmetric Lagrangian with pseudoscalar, 
vector and axial vector mesons \cite{mei88}, we start with the non-linear $\sigma$ model 
\begin{equation}
\mathcal{L} = \frac{1}{8}F_\pi^{\,2}{\rm {\rm Tr}}\left(\partial_\mu U 
\partial^{\mu}U^{\dagger}\right),
\label{nsigma}
\end{equation}
where $U = \exp(2i\phi/F_\pi)$ and $\phi = \frac{T^a\phi^a}{\sqrt 2}$, $T^a$ are the $SU(N_f)$ generators, and 
we introduce the vector and axial vector mesons by minimal coupling of the right- and 
left-handed vector fields
\begin{eqnarray}
A_{L\mu} = \frac{1}{2}\left(V_{\mu}+A_{\mu}\right) \\
A_{R\mu} = \frac{1}{2}\left(V_{\mu}-A_{\mu} \right).
\end{eqnarray}
The resulting Lagrangian
\begin{equation}
\mathcal{L} = \frac{1}{8}F_\pi^{\/2}{\rm Tr}\left[D_\mu U 
D^{\mu}U^{\dagger}\right] 
-\frac{1}{2}{\rm {\rm Tr}}\left[F_{\mu\nu L}F^{\mu\nu}_L +  
F_{\mu\nu R}F^{\mu\nu}_R \right],
\label{LGAUGE}
\end{equation}
where $D_\mu U = \partial_{\mu}U-igA^L_{\mu}U+igUA^R_{\mu}$, $F^{R,L}_{\mu\nu}$ 
are the non-Abelian field strength tensors, and $g$ is the universal gauge 
coupling, is then invariant under a $SU_L(N_f)\times SU_R(N_f)$ 
transformation\cite{kay85}
\begin{eqnarray}
U &\rightarrow& U_LUU^{\dagger}_R \label{trans1}\\
A_\mu^{L} &\rightarrow& U_L A^L_\mu U^{\dagger}_L + \frac{i}{g}U_L\partial_\mu U^{\dagger}_L \label{trans2}\\
A_\mu^{R} &\rightarrow& U_R A^R_\mu U^{\dagger}_R + \frac{i}{g}U_R\partial_\mu 
U^{\dagger}_R.
\label{trans3}
\end{eqnarray}
We note here that $SU_V(N_f)$ and $SU_A(N_f)$ are subgroups of the full $SU_L(N_f)\times SU_R(N_f)$ group. To account for the right- and left-handed mesons' masses, we add the 
symmetry-invariant term
\begin{equation}
m_0^{\/2}{\rm Tr}\left[A_{\mu L}A^{\mu}_L+A_{\mu R}A^{\mu}_R\right].
\label{LMASS1}
\end{equation}
We could also supplement the Lagrangian with the pseudoscalar mass term \cite{mei88}
\begin{equation}
\frac{1}{8}F_\pi^{\,2} {\rm Tr}\left(M(U+ U^\dagger -2)\right),
\end{equation} 
where $M$ is the pseudoscalar mass matrix. But this term explicitly breaks the 
symmetry, and thus will be considered as a correction (as lifting the mass 
degeneracy of the vector mesons). Expanding the Lagrangian and removing the 
mixing between the pseudoscalar and axial vector fields via 
\begin{eqnarray}
A_{\mu} \rightarrow A_\mu + \frac{g\tilde F_{\pi}}{2m_0^2}\partial_{\mu}\phi ,\mbox{  } \phi \rightarrow Z^{-1}\tilde \phi , \mbox{  } F_\pi \rightarrow Z^{-1}\tilde F_\pi, \mbox{  }Z^2 = \left(1-\frac{g^2\tilde F_\pi^{\,2}}{4m_0^2}\right)
\end{eqnarray}
yields
\begin{eqnarray}
\mathcal{L} &=& \frac{1}{2}{\rm Tr}\left[\partial_\mu\phi\partial^{\mu}\phi\right] -\frac{1}{4}{\rm Tr}\left[F^V_{\mu\nu }F^{\mu\nu}_V +  F^A_{\mu\nu}F^{\mu\nu}_A \right] \nonumber  \\
&+&\frac{1}{2}m_V^2{\rm Tr}\left[V_\mu^2\right]
+\frac{1}{2}m_A^2{\rm Tr}\left[A_\mu^2\right] \nonumber \\
&+&\mathcal{L}_{V\phi\phi} + \mathcal{L}_{AV\phi} + \mathcal{L}_{VV\phi\phi} + \cdots\,,
\label{LVAP}
\end{eqnarray}
where the tildes were dropped for simplicity and the degenerate masses are 
defined as $m_V^2 = m_0^2$, and $m_A^2 = m_V^2/Z^2$. To the above Lagrangian 
other non-minimal terms can be added and are necessary to fit $\pi$, 
$\rho$, and $a_1$ phenomenology~\cite{gom84}. But the Lagrangian of 
Eq. (\ref{LVAP}) will be sufficient for our purposes.


\subsection{$SU_L(N_f)\boldmath{\mbox{$\times$}} SU_R(N_f)$ Lagrangian with 
pseudoscalar and vector mesons }


In the previous section the axial mesons were introduced as the
chiral partners of the $\rho$ fields 
resulting in a linear realization of the symmetry \cite{gai69}. In the present 
case, since the desired $SU_L(N_f)\times SU_R(N_f)$  Lagrangian will involve 
only pseudoscalar and vector mesons, both will then have to transform 
non-linearly under the symmetry. This is similar to building an effective 
low-energy theory with the $\pi$ field transforming non-linearly under 
the axial group, or equivalently by imposing a  constraint \cite{gai69,wei68}. 
The second approach will be favoured here by gauging-away the axial 
mesons~\cite{kay85}. 
But before doing so, we add to Eq.~(\ref{LGAUGE}) the locally-invariant term 
\begin{eqnarray}
\gamma {\rm Tr}\left(F_{\mu\nu}^LUF^{\mu\nu R}U^{\dagger}\right)
\end{eqnarray}
and the mass term of Eq.~(\ref{LMASS1}) with the further addition
\begin{eqnarray}
B{\rm Tr}\left[A_{L\mu}UA^\mu_RU^\dagger\right].
\end{eqnarray}
The second mass term is non-minimal and is introduced to account for 
$\rho$-meson phenomenology in the final Lagrangian \cite{kay85}. To remove 
the axial mesons we set $U_L = U^{1/2}=\zeta$ and $U_R = U^{-1/2} = 
\zeta^{\dagger}$ in Eqs. (\ref{trans1}-\ref{trans3}) giving
\begin{eqnarray}
U &=& \zeta1\zeta \\
A_\mu^{L} &=& \zeta\rho_\mu\zeta^{\dagger} + \frac{i}{g}\zeta\partial_{\mu}\zeta^{\dagger} \\
A_\mu^{R} &=& \zeta^{\dagger}\rho_\mu\zeta + \frac{i}{g}\zeta^{\dagger}\partial_{\mu}\zeta\,.
\end{eqnarray}
This amounts to imposing the $SU_L(N_f)\times SU_R(N_f)$ invariant constraint 
\begin{equation}
\mathcal{D}_\mu U^\dagger = 0.
\end{equation}
The new vector field $\rho_\mu$ then transforms in the usual way under the 
vector symmetry, namely  
\begin{eqnarray}
\rho_\mu \rightarrow K\rho_\mu K^\dagger + \frac{i}{g}K\partial_\mu K^\dagger,
\end{eqnarray}
where $K \in SU_V(N_f)$, but transforms non-linearly under the axial-vector
subgroup. 
Substituting the new field definitions and normalizing the vector and 
pseudoscalar kinetic terms by choosing 
\begin{eqnarray}
\gamma &=& \frac{3}{4}, \mbox{ } \frac{2m_0^2-B}{g^2F_\pi^{\,2}} = \frac{1}{2}
\end{eqnarray}
yields 
\begin{eqnarray}
\mathcal{L} &=& -\frac{1}{4}{\rm Tr}\left[F_{\mu\nu}(\rho)F^{\mu\nu}(\rho)\right] + \frac{1}{2}m_V^2{\rm Tr}[\rho_\mu^2] \nonumber \\
&+&\frac{F_\pi^{\,2}}{4}(1+k){\rm Tr}\left[\partial_{\mu}\zeta\partial^{\mu}\zeta\right]+ i\frac{F_\pi^{\,2}\,g_{V\phi\phi}}{2}{\rm Tr}\left[\rho^\mu\left(\partial_\mu\zeta\zeta^\dagger + \partial_\mu\zeta^\dagger\zeta\right)\right] \nonumber \\
&+&\frac{F_\pi^{\,2}}{4}(1-k){\rm Tr}\left[\zeta^\dagger\partial^\mu\zeta^\dagger\zeta\partial_\mu\zeta\right],
\label{LVP1}
\end{eqnarray}
where the parameters are defined as $m_V^2 = 2B+4m_0^2$, $g_{V\phi\phi}= 
m_V^2/gF_\pi^{\,2}$ and $g= g_{V\phi\phi}/k$, and the field strength tensor is 
$F_{\mu\nu}(\rho) = \partial_\mu \rho_\nu - \partial_\nu \rho_\mu -ig[\rho_\mu,
\rho_\nu]$ \cite{sch86}. Notice the extra parameter $k$, owing to the 
introduction of the non-minimal vector meson mass term. As with the 
Lagrangian 
of Eq.~(\ref{LVAP}), this Lagrangian of pseudoscalars and vector mesons is 
globally invariant under $SU_L(N_f)\times SU_R(N_f)$, and consequently under 
the $SU_V(N_f)$ subgroup.



\subsection{$SU_V(N_f)\boldmath{\mbox{$\,$}}$ Lagrangian with 
pseudoscalar and vector mesons }


The two preceding Lagrangians encode $SU_L(N_f)\times SU_R(N_f)$ symmetry.  For
the purpose of reviewing the effective Lagrangian models used to
calculate charmonium dissociation, we now build a $SU_V(N_f)$ invariant-only 
Lagrangian by gauging 
the non-linear $\sigma$ model with the vector meson field \cite{sak69}
\begin{eqnarray}
\mathcal{L} = \frac{F_\pi^{\,2}}{8}{\rm Tr}\left[D_\mu UD^\mu U^\dagger\right] -
\frac{1}{4}{\rm Tr}[F_V^{\mu\nu}F^V_{\mu\nu}] &+&\frac{1}{2}m_V^2{\rm Tr}
[V_\mu^2],
\label{LVP2}
\end{eqnarray}
where $D_\mu U = \partial_\mu U -i\frac{g}{2}[V_\mu,U]$ and $F^V_{\mu\nu} = 
\partial_\mu V_\nu -\partial_\nu V_\mu - i\frac{g}{2}[V_\mu,V_\nu]$. This 
Lagrangian is invariant under a global $SU_V(N_f)$ transformation, but not 
under the full $SU_L(N_f)\times SU_R(N_f)$ symmetry group ($SU_V(N_f)$ is a subgroup of this group). It also exhibit a 
$PPVV$ contact interaction, while Eq.~(\ref{LVP1}) has no such term.


\section{Adler's Theorem}


A spontaneously broken symmetry not only implies the existence of Goldstone 
bosons, but also constrains their low-energy behaviour. Here we will consider 
the transition amplitude for emitting one Goldstone boson (the proof can 
easily be extended to more than one). Following Weinberg \cite{wei95}, we first 
note that the current operator can create a Goldstone boson
\begin{equation}
\left<0|J^\mu(x)|B\right> = i\frac{Fq^\mu}{(2\pi)^3 2E} e^{-iq\cdot x}
\end{equation}
where F is the decay constant. We expect the matrix element $\left<\beta|J^\mu(x)|\alpha\right>$ to
then have a pole term. Indeed, in general, we can write
\begin{equation}
 \left<\alpha|J^\mu(0)|\beta\right> = N^\mu_{\beta\alpha} + i\frac{F q^\mu}{q^2}
 M^B_{\beta \alpha},
\end{equation} 
where $M^B_{\beta \alpha}$ is the desired transition amplitude for emitting one 
Goldstone boson, and $N^\mu_{\beta \alpha}$ is assumed to be the pole-free 
contribution. Applying the current conservation constraint, we then find that
\begin{equation}
M^B_{\beta \alpha} = \frac{i}{F}q_\mu N^\mu_{\beta \alpha} 
\end{equation}
and we see that as $q \rightarrow 0$ the RHS vanishes. Historically, this 
property was first studied by Adler \cite{adl65}, and rests on the assumption
that $N^\mu_{\beta\alpha}$ has no singularity as $ q\rightarrow 0$. This is not
valid in general, since a pole may arise through the insertion of the current 
operator on an external line \cite{sak69}. $M^B_{\beta \alpha}$ can then have 
a non-vanishing contribution in the soft limit due to emission of a soft 
Goldstone boson off an external line.
But this is not the case for $SU_L(N_f) \times SU_R(N_f)$ normal parity mesonic 
Lagrangians 
where there are no vertices such as $\phi^3$ or $VV\phi$ (see Eq.~\ref{LVP1}).
\\\indent Thus, for the normal parity 
Lagrangians where the underlying symmetry is $SU_L(N_f)\times SU_R(N_f)$, we 
expect that transition amplitudes will vanish in the pseudoscalar zero-momentum  
limit \cite{bur00}, up to corrections from coupling to other non-symmetric 
invariant gauge fields, such as the electromagnetic field, and in general any 
non-invariant terms (e.g. pseudoscalar masses).  In the following subsections, we 
will explicitly check that decoupling occurs for the process
$\pi^+ + \rho^0 \rightarrow \pi^+ + \rho^0$ in the two chiral-invariant
Lagrangians of Eqs.~(\ref{LVAP}-\ref{LVP1}), but not for the isospin-invariant 
model of Eq.~(\ref{LVP2}).
 

\subsection{Chiral model with $\mbox{\boldmath{$\pi$}}$, 
$\mbox{\boldmath{$\rho$}}$, and $\boldmath{\mbox{$a_1$}}$ mesons}

\begin{figure}[!h]
\begin{center}
\includegraphics[scale=0.75]{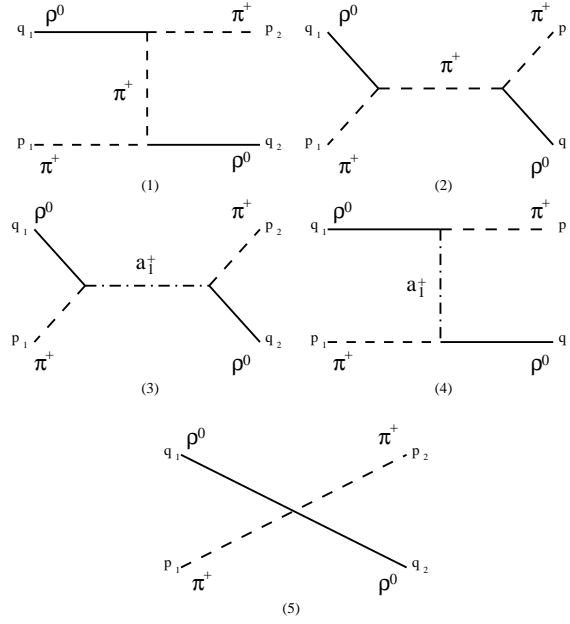}
\caption{$\rho^0 + \pi^+ \rightarrow \rho^0 + \pi^+$.}
\label{fig1}
\end{center}
\end{figure}
The process we want to consider, namely $\pi^+(p_1)+\rho^0(q_1) \rightarrow 
\pi^+(p_2)+\rho^0(q_2)$, involves 5 tree-level diagrams for the given particle 
content: $\pi$- and $a_1$-mediated exchanges both in $s$ and $t$ channels, and 
through a 4-point interaction (Fig~\ref{fig1}). We start with the relevant 
interaction terms from the Lagrangian of Eq.~(\ref{LVAP})
\begin{eqnarray}
\mathcal{L}_{V\phi\phi} &=& -\frac{ig}{2}{\rm Tr}\Big[V^\mu[\phi,\partial_\mu 
\phi]\Big] +\frac{igZ^2}{2m_V^2}{\rm Tr}\Big[(\partial_\mu V_\nu - \partial_\nu 
V_\mu)\partial^\mu \phi \partial^\nu \phi\Big] \\
\mathcal{L}_{AV\phi}&=& \frac{ig^2\,F_\pi}{4m_V^2}{\rm Tr}\Big[(\partial_\mu 
V_\nu-\partial_{\nu} V_\mu)[A^\mu,\partial^\nu\phi] + (\partial_\mu A_\nu-
\partial_{\nu} A_\mu)[V^\mu,\partial^\nu \phi]\Big] \Big] \nonumber \\ 
&-&\frac{ig^2F_\pi}{4Z^2}{\rm Tr}\Big[V^\mu[A_\mu,\phi]\Big]\\
\mathcal{L}_{VV\phi\phi} &=& \frac{g^2}{8}\frac{g^2F_\pi^{\,2}}{4m_0^4}{\rm Tr}
\Big[[V^\mu,V^\nu][\partial_\mu \phi,\partial_\nu \phi] + [V^\mu,
\partial^\nu \phi][\partial_\mu \phi,V_\nu]+ [V^\mu,\partial^\nu\phi]
[V_\mu,\partial_\nu \phi]\Big] \nonumber \\
&-&\frac{g^2}{8Z^2}{\rm Tr}\Big[[V_\mu,\phi]^2\Big].
\end{eqnarray}
From here we extract the off-shell vertex functions
\begin{eqnarray}
\Gamma^1_{\mu} &=& -\frac{g}{\sqrt{2}}\left[k_\mu+ p_\mu -\frac{(1-Z^2)}
{m_V^2}\left(q\cdot k p_\mu-q\cdot p k_\mu\right)\right] \\
\Gamma^{2}_{\mu\nu} &=& -\frac{ig}{\sqrt{2}}\frac{(1-Z^2)^{\frac{1}{2}}}{m_V}
\left\{(m_A^2+q^2-k^2)g_{\mu\nu} -q_\mu q_\nu+k_\mu k_\nu\right \} \\
\Gamma^{3}_{\mu\nu} &=& 2\left[\frac{g^2}{2}\frac{(1-Z^2)}{2m_V^2}\left\{-2p_1 
\cdot p_2 g_{\mu\nu} +
p_{1\mu}p_{2\nu} +p_{2\mu}p_{1\nu}\right\} + \frac{g^2}{2Z^2}g_{\mu\nu} \right]
\end{eqnarray}
for $\pi^+(p)+\rho^0(q) \rightarrow \pi^+(k)$, $\pi^+(p)+\rho^0(q) \rightarrow 
a_1^{+}(k)$, and the four-point interaction, respectively. The resulting five 
off-shell amplitudes are then
\begin{eqnarray}
i\mathcal{M}^1_{\mu\nu} &=& i\Gamma^{1 \dagger}_{\mu}\frac{i}{s-m_\pi^2}
i\Gamma^{1}_\nu \\
i\mathcal{M}^2_{\mu\nu} &=& i\Gamma^{1 \dagger}_{\mu}\frac{i}{t-m_\pi^2}
i\Gamma^{1}_\nu \\
i\mathcal{M}^3_{\mu\nu} &=&
 i\Gamma^{2 \dagger}_{\mu\alpha}\frac{-i[g^{\alpha\beta}-
 (p_1+q_1)^\alpha(p_1+q_1)^\beta/m_A^2]}{s-m_A^2}i\Gamma^{2}_{\nu\beta}
\\
i\mathcal{M}^4_{\mu\nu}&=& i\Gamma^{2 \dagger}_{\mu\alpha}\frac{-i
[g^{\alpha\beta}-(q_2-p_1)^\alpha(q_2-p_1)^\beta/m_A^2]}{t-m_A^2}
i\Gamma^{2}_{\nu\beta}\\
i\mathcal{M}^5_{\mu\nu} &=& i\Gamma^{3}_{\mu\nu}.
\end{eqnarray}
Having now the full amplitude for the given process, we wish to see if the 
pseudoscalar decoupling theorem holds. The presence of the contact term in the 
4-point interaction leads us to expect cancellations to occur as we let one
of the 
pions' 4-momentum go to zero. Stated differently, since the $\rho$-meson 
was introduced in a chirally symmetric way by adding its chiral partner 
(i.e. $a_1$), 
the transition amplitude relies on help from the $a_1$ in what
amounts to a delicate cancellation allowing the pion to
decouple.  To show 
this, we contract the amplitudes with the appropriate polarization vectors, 
and then let $p_2 \rightarrow 0$ (a similar proof can be shown to hold for $p_1 
\rightarrow 0$). First it is seen that the amplitudes involving the 
$\pi$-exchange 
go to zero because of transversality (i.e $\epsilon(q)\cdot q$).  For 
$a_1$-exchange in the $s$-channel we find
(the same result is true for the $t$-channel) 
\begin{eqnarray}
i\mathcal{M}_3 &=& \epsilon^{*\mu}(q_2)
\left[i\Gamma^{2\dagger}_{\mu\alpha}(p_2 \rightarrow 0) \frac{-i
\left[g^{\alpha\beta}-\frac{q_2^{\alpha}q_2^{\beta}}{m_A^2}\right]}
{m_V^2-m_A^2}i\Gamma^{2}_{\nu\beta}\right]\epsilon^{\nu}(q_1) \nonumber \\
&=& -\frac{ig^2}{2}\frac{1}{m_V^2}\epsilon^{*\mu}(q_2)\left[
m_A^2 g_{\mu\nu}-q_{1\mu}q_{1\nu} +\frac{q_1 \cdot q_2}{m_A^2}
q_{2\mu}q_{1\nu} + \frac{m_V^2}{m_A^2}q_{2\mu}q_{2\nu}\right]
\epsilon^{\nu}(q_1) \nonumber \\
&=& -\frac{ig^2}{2Z^2}\epsilon^*(q_2)\cdot \epsilon(q_1).
\end{eqnarray}
The last line comes about again due to the orthogonality condition. Finally, 
the 4-point interaction reads
\begin{equation}
i\mathcal{M}_5 = \frac{ig^2}{Z^2}\epsilon^*(q_2)\cdot \epsilon(q_1),
\end{equation}
and thus the {\it full amplitude} is shown to vanish as expected. Note the 
cancellation between the 4-point interaction and the $a_1$ channels. In summary,
even though the pions are not coupled through gradient coupling for all 
interaction terms, the net amplitude still vanishes due to intricate cancellations
amongst all the channels.


\subsection{Chiral model with $\boldmath{\mbox{$\pi$}}$ and 
$\boldmath{\mbox{$\rho$}}$ mesons }


In this model there are only two diagrams, namely $s\,$- and $t\,$-channels of 
pion exchange. The relevant interaction Lagrangian is
\begin{equation}
\mathcal{L}_{\rho\pi\pi} = -i\frac{g_{\rho\pi\pi}}{2}{\rm Tr}
\left[\rho^\mu[\phi,\partial_\mu \phi]\right]\,
\end{equation}
and the extracted vertex for the s- and t- channels is
\begin{eqnarray}
\Gamma_\mu^{'1} = -\frac{g_{\rho\pi\pi}}{\sqrt 2}\left(p_\mu+k_\mu\right), \\
\end{eqnarray}
with $k=p+q$ and $k=p-q$, respectively. The two amplitudes are then
\begin{eqnarray}
i\mathcal{M}^1_{\mu\nu} = i\Gamma^{'1 \dagger}_{\mu}\frac{i}{s-m_\pi^2}i\Gamma^{'1}_\nu \\
i\mathcal{M}^2_{\mu\nu} = i\Gamma^{'1 \dagger}_{\mu}\frac{i}{t-m_\pi^2}i\Gamma^{'1}_\nu\,.
\end{eqnarray}
We immediately see that the soft pion theorem holds separately for each 
amplitude when the proper polarization vectors are contracted with the 
amplitudes.


\subsection{Isospin-invariant model with $\boldmath{\mbox{$\pi$}}$ and 
$\boldmath{\mbox{$\rho$}}$}

Here the interaction Lagrangians are given by
\begin{eqnarray}
\mathcal{L}_{\rho\pi\pi} &=&-i\frac{g}{2}{\rm Tr}\Big[V^\mu[\phi,\partial_\mu \phi]\Big] \\
\mathcal{L}_{\rho\rho\pi\pi} &=& -\frac{g^2}{8}{\rm Tr}\Big[[V_\mu,\phi]^2\Big].
\end{eqnarray}
We note that the three-point interaction is identical in structure to the ones
from the previous sections, and the derived amplitude from these
terms disappears in the appropriate limit. The difference lies here in the 
additional 4-point interaction. Indeed, in the zero-momentum limit for the
pion its contribution is non-vanishing, and therefore the 
pions do not decouple. It is then expected that the behaviour of the
cross-section 
for the process near threshold to be different from the one calculated
in the chiral models. 


\section{ELECTROMAGNETIC CURRENT CONSERVATION}

We now wish to add electromagnetism to all three models and to investigate the
validity of vector meson dominance (VMD) \cite{sak69}. 

\subsection{$SU_L(N_f)\boldmath{\mbox{$\times$}} SU_R(N_f)$ Lagrangian with 
pseudoscalar, vector and 
axial vector mesons}

The fields transform under $U_{EM}(1)$ as \cite{sch86}:
\begin{eqnarray}
\delta a_\mu &=& \frac{1}{e}\partial_\mu\epsilon \\
\delta U &=& i\epsilon[Q,U] \\
\delta A_\mu^{L,R} &=& i\epsilon[Q,A_\mu^{L,R}]+\frac{1}{g}Q
\partial_\mu\epsilon,
\end{eqnarray}
where $a_\mu$ is the electromagnetic field and $Q$ is the appropriate quark 
charge matrix. Using Witten's iterative method \cite{wit83}, 
Eq.~(\ref{LVAP}) can be made invariant provided we add the following terms 
\cite{sch86}
\begin{eqnarray}
\Delta_{L} = -\frac{2em_0^2}{g}a^\mu {\rm Tr}\left[Q(A_\mu^L + A_\mu^R)\right]
 +\frac{2em_0^2}{g^2}a_\mu^2 {\rm Tr}Q^2\,.
\label{EM1}
\end{eqnarray}
By explicitly expanding the first term one obtains \cite{kro67}
\begin{equation}
\mathcal{L}_{\rm VMD} = -\sqrt 2\frac{e}{g}m_V^2a^\mu\rho^0_\mu + \cdots 
\,.
\end{equation}
The above interaction term is precisely Sakurai's original formulation of  
vector meson dominance.

\subsection{$SU_L(N_f)\boldmath{\mbox{$\times$}} SU_R(N_f)$ Lagrangian with
pseudoscalar and vector mesons}

We now turn to the non-linear model. Here we must add extra-terms due to the 
non-minimal mass term introduced in the mesonic Lagrangian of Eq.~(\ref{LVP1}). 
The counter terms added to make the Lagrangian invariant are
\begin{eqnarray}
\Delta_{NL} &=& -\frac{2em_0^2}{g}a^\mu {\rm Tr}\left[Q(A_\mu^L + A_\mu^R)\right]
 - \frac{B}{g}a^\mu {\rm Tr}\left[Q\left(UA_\mu^RU^\dagger + U^\dagger 
 A_\mu^LU\right)\right]\nonumber \\ &+&\frac{2em_0^2}{g^2}a_\mu^2 
 {\rm Tr}Q^2+\frac{Be^2}{g^2}a_\mu^2{\rm Tr}\left[QUQU^\dagger\right].
\end{eqnarray}
Then using the expressions for $A_\mu^L$, $A_\mu^R$ and $U$ in terms of $\rho$ 
and $\phi$, we find
\begin{eqnarray}
\Delta_{NL} &=& -ea^\mu\left[kgF_\pi^{\,2}{\rm Tr}[Q\rho_\mu] +i(1-k/2){\rm Tr}
\left[Q[\phi,\partial_\mu \phi]\right] - \frac{g_{\rho\pi\pi}}{2}{\rm Tr}
\left[Q\left[\phi,[\phi,\rho_\mu]\right]\right]\right] + \cdots\,.
\end{eqnarray}
We first note that beyond the vector-photon coupling (first term), there can 
be a direct $\phi\phi\gamma$ contribution (second term). Moreover a 4-point 
interaction exists (third term) of type $V\phi\phi\gamma$ and is essential for 
current conservation in processes such as $\rho^0 + \pi^+ \rightarrow  
\pi^+ + \gamma$.    
 
\subsection{$SU_V(N_f)\boldmath{\mbox{$\,$}}$ Lagrangian with pseudoscalar
and vector mesons }
Finally, we examine the $SU_V(N_f)$-invariant Lagrangian of Eq.~(\ref{LVP2}). 
Under the electromagnetic gauge transformations, since the photon couples only to 
the vector field, the counter terms are also given by $\Delta_{L}$
(Eq.~\ref{EM1}).


\section{CHARMONIUM EFFECTIVE LAGRANGIANS}


We now review the effective Lagrangians used to calculate the $J/\psi$ 
dissociation rate in a hadronic gas \cite{hag00, lin00, oh01}. The underlying
hypothesis of all these models is to assume that the pseudoscalar and vector
meson fields are in multiplets of $SU(4)$. The interactions are then built 
from these by using the same techniques outlined in previous sections. 
Coupling constants for the various interactions are fixed either 
empirically or using symmetry arguments \cite{hag00, lin00, oh01}. Moreover,
to model short range interactions, 
form factors are introduced (e.g. \cite{hag00}). Overall, the models differ in 
their methods for fixing the coupling constants, their
choice of form factors and implementation, and in their abnormal
parity interaction content. \\\indent The three approaches can be summarized 
as Lagrangians where 
(i) both axial vector and vector mesons are present, but the interaction 
vertices with 
axial vector mesons are dropped \cite{oh01}, (ii) non-minimal mass terms are added and 
the axial mesons are gauged-away by imposing a $SU_L(4)\times SU_R(4)$ symmetric 
constraint in a similar fashion as in Section II. C \cite{hag00}, and (iii) only 
the vector mesons are introduced through gauge-coupling as in section II. B 
\cite{lin00}. Comparing the Lagrangians in references \cite{hag00, lin00, oh01}, 
we see that, up to a constant redefinition, they all lead to the same normal 
parity  $SU_V(4)$ invariant-only Lagrangian, namely that of Eq.~(\ref{LVP2}) 
\cite{not01}. Indeed, all these approaches lead to the pseudoscalars not 
decoupling for the $P + V \rightarrow P + V$  process in the zero-momentum 
pseudoscalar limit for degenerate vector mesons masses. In the first approach, 
the theorem is not respected
because the axial vector mesons are omitted, and as it was shown 
that virtual axial vector meson
exchange plays an essential role in canceling the non-vanishing zero-momentum 
contribution of the $PPVV$ vertex. In the second approach,
the theorem was again not respected because the third
non-minimal mass term of Eq.~(2.6) in \cite{hag00} is not globally invariant 
under $SU_L(4)\times SU_R(4)$ symmetry, which leads to a non-vanishing
zero-momentum $PPVV$ contribution. And in the third case, the Lagrangian is globally invariant only under $SU_V(4)$, but not the extended symmetry \cite{lin002}. Thus, none of the amplitudes extracted
from these Lagrangians obeys the decoupling theorem in the degenerate vector
mass limit, neither for the 
$J/\psi + \pi \rightarrow D^{*}(\bar D^*) + \bar D(D)$ nor the {$\rho + \pi 
\rightarrow \rho + \pi$ processes.


\section{Comparative ANALYSIS}


We have shown that the effective Lagrangian models found in the literature 
\cite{hag00, lin00, oh01} are all $SU_V(4)$ invariant (in the degenerate mass 
limit), but not  $SU_L(4) \times SU_R(4)$ invariant. Here we will compare the 
$J/\psi + \pi$ cross-section as calculated within the models of Eq.~(\ref{LVP1}) 
and Eq.~(\ref{LVP2}). In the former, we need to consider three diagrams 
(Fig.~\ref{fig2}), while for the latter, the contact interaction is absent (as 
expected from previous discussions). In Appendix A, we further address the 
pseudoscalars' decoupling in the soft momentum limit, and in Appendix B we 
investigate the  electromagnetic current conservation for the related $\gamma 
+ \pi \rightarrow \bar{D} + D^*$ process.

\begin{figure}[!h]
\begin{center}
\includegraphics[scale=0.75]{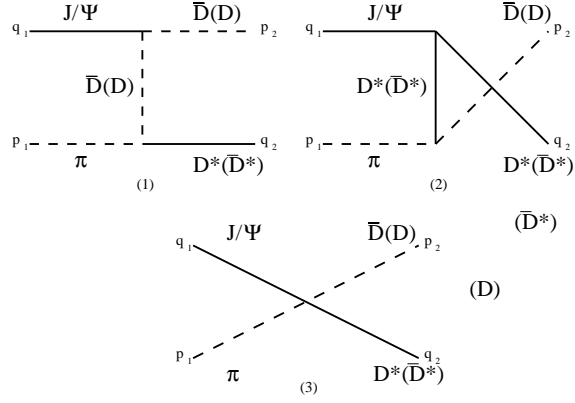}
\caption{$J/
\psi + \pi \rightarrow D^{*}(\bar D^{*}) + \bar D(D)$.}
\label{fig2}
\end{center}

\end{figure}
 

\subsection{$\boldmath{\mbox{$SU_V(4)$}}$ model}


Defining the pseudoscalar and vector field matrices as in \cite{hag00}, the 
relevant 
interaction terms from Eq.~(\ref{LVP2}) are 
\begin{eqnarray}
\mathcal{L}_{J/\psi DD} &=& ig_{J/\psi DD} \psi^\mu(D\partial_\mu\bar D - 
\partial_\mu D \bar D) \label{int1} \\
\mathcal{L}_{\pi DD^*} &=& ig_{\pi DD^*} D^{*\mu}(\partial_\mu \pi \bar D - 
\pi \partial_\mu \bar D) + h.c. \\
\mathcal{L}_{J/\psi D^* D^*} &=& ig_{J/\psi D^*D^*}[\psi^\mu(\partial_\mu 
D^{*}_\nu \bar D^{*\nu} -
D^{*\nu}\partial_{\mu}\bar D^{*}_\nu) + \psi^\mu(D^{*\nu}\partial_\nu \bar 
D^*_\mu - \partial_{\mu}D^{*}_\nu \bar D^{*\nu}) \nonumber \\
&-&\partial_\mu \psi_\nu (D^{*\mu}\bar D^{*\nu}-D^{*\nu}\bar D^{*\mu})] 
\label{int3} \\
\mathcal{L}_{J/\psi D^*D\pi} &=& -g_{J/\psi D^*D\pi}\psi^\mu (D^*_\mu \pi 
\bar D + D \pi \bar D^*_\mu),
\end{eqnarray}
where $\psi$ is the $J/\psi$ field, the isospin doublets are \cite{oh01}
$\bar D^T = (\bar D^0,
D^-)$,  $D = (D^0, D^+)$, and similarly for the vector particles, and $\pi = 
\frac{1}{\sqrt 2} \vec \tau \cdot\vec \pi$. Provided the symmetry is exact, we 
have also
\begin{eqnarray}
g_{J/\psi DD} =g_{J/\psi D^*D^*}  = \frac{g}{\sqrt 3}, \mbox{ } g_{\pi DD^*} = 
\frac{g}{2}, \mbox{ }
g_{J/\psi D^*D\pi} = \frac{g^2}{2\sqrt 3}\,.
\end{eqnarray}
The amplitudes for the absorption process are then
\begin{eqnarray}
i\mathcal{M}^1_{\mu\nu} &=&  +ig_{J/\psi DD}g_{\pi DD^*}\frac{(2p_{2\nu}-q_{1\nu})
(2p_{1\mu}-q_{2\mu})}{(q_1-p_2)^2-m_{D}^2}  \label{amp1}\\
i\mathcal{M}^2_{\mu\nu} &=& -ig_{J/\psi D^*D^*}g_{\pi DD^*}\frac{[(2q_2-q_1)_\nu 
g_{\mu\lambda} + (2q_1-q_2)_\mu g_{\lambda\nu} -(q_1+q_2)_\lambda g_{\mu\nu}]
(p_{1\lambda}(1+\Delta)+p_{2\lambda}(1-\Delta))}{(q_1-q_2)^2-m_D^{*2}} \label{amp2}\\
i\mathcal{M}^{3}_{\mu\nu} &=& -ig_{J/\psi D^*D\pi}g_{\mu\nu},
\end{eqnarray}
where $\Delta = (m_D^2-m_\pi^2)/m_{D^*}^2$. In Appendix A we show that for 
degenerate vector meson masses, the full amplitude does not vanish in the 
soft-momentum limit. Rather, there is a left-over contact term due to the 
third diagram.


\subsection{$SU_L(4)\boldmath{\mbox{$\times$}} SU_R(4)$ model}


For the $SU_L(4) \times SU_R(4)$ invariant Lagrangian of Eq.~(\ref{LVP1}), the 
interaction terms are given by Eq~(\ref{int1})-(\ref{int3}), but now with  
\begin{eqnarray}
g_{J/\psi DD} = \frac{g_{V\phi\phi}}{\sqrt 3} =\frac{kg}{\sqrt 3},\mbox{ }  
g_{J/\psi D^*D^*} = \frac{2g}{\sqrt 3}, \mbox{ } g_{\pi DD^*} = \frac{g_{V\phi\phi}}{2} 
= \frac{kg}{2}.
\end{eqnarray}
With these, the two relevant amplitudes are those of Eqs.~(\ref{amp1}) and 
(\ref{amp2}). 
Note that we can map the $SU_V(4)$ couplings to those above by letting $k=2$ and 
setting $g \rightarrow g/2$.


\subsection{Results}


The first step is to fix the coupling constants of the two models. This is done
in Appendix C. Also, to account for short range interactions form 
factors would have to be folded in \cite{hag00, lin00, oh01}.
But, since we are here
interested in the effect of the implementation of the symmetry group, they 
will not be introduced.  The differential isospin-averaged cross-section 
is then given by
\begin{equation}
\frac{d\sigma}{dt} = \frac{1}{128\pi s p^2_{1}} 
\mathcal{M}^{\mu\nu}\mathcal{M}^{\alpha\beta}\left[g_{\mu\alpha}-
\frac{q_{2\mu}q_{2\alpha}}{m^2_{D^*}}\right]\left[g_{\nu\beta}-\frac{q_{1\nu}q_{1\beta}}
{m^2_{J/\psi}}\right],
\end{equation} 
where the appropriate model-dependent squared amplitude is used, an
isospin factor 
of two has been included and the centre of mass momentum is 
\begin{equation}
p^2_{1} = \frac{1}{4s}\lambda(s,m^2_\pi,m^2_{J/\psi})
\end{equation}
and the triangle function is $\lambda(x,y,z) = x^2 -2x(y+z)+(y-z)^2$.
Integrating over the kinematical range defined by
\begin{equation}
t_\pm = m_\pi^2 + m_{D^*}^2 - \frac{1}{2s}(s+m_\pi^2 - m_{J/\psi}^2)(s+m_{D^*}^2 
- m_{D}^2) \pm \frac{1}{2s}\lambda^{1/2}(s,m_\pi^2,m_{J/\psi}^2)\lambda^{1/2}
(s,m_D^2,m_{D^*}^2)
\end{equation}
gives the total cross-section. Carrying this to completion 
for the two models yields Fig.~\ref{fig3}. 
We see an energy-dependent reduction in the cross sections across 
the relevant domain and to quote a specific number we note
that at $\sqrt s = 5$ GeV the cross-section is reduced by 
about 40\% going from the $SU_V(4)$ model to the $SU_L(4) \times SU_R(4)$ model.   

\begin{figure}[!h]
\begin{center}
\includegraphics[scale=0.50]{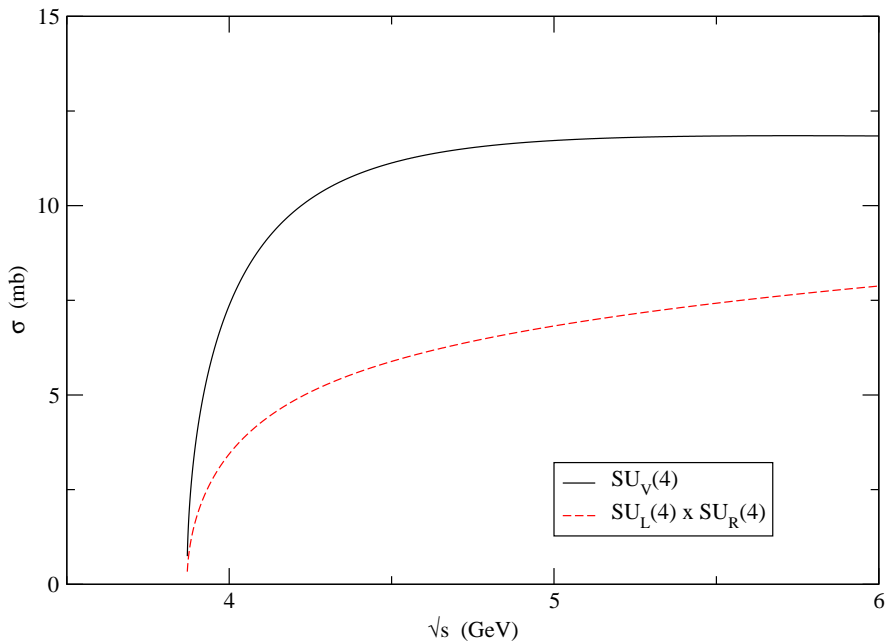}
\caption{Isospin-averaged cross-section for $J/\psi + \pi \rightarrow (D^* + 
\bar D) + (\bar D^* + D) $.}
\label{fig3}
\end{center}
\end{figure}


\section{CONCLUSION AND OUTLOOK}

Modeling low-energy hadron physics is particularly challenging
when there is limited experimental input available
for constraints.  This is the current situation in the 
charm sector as the only measurement relevant for fixing
coupling constants in the model is the
decay width for $D^{*}\rightarrow\/D\pi$~\cite{cleo}.
We have therefore invoked symmetries and general theorems.
In particular, we have checked for full $SU_L(4)\times\/SU_R(4)$
symmetry and the appropriate limit to test for compliance with Adler's 
theorem. We found that none of the published models can 
do this, and we therefore proposed a new effective Lagrangian---the
first one which does encode complete four-flavour chiral symmetry and Adler's theorem.
Our interest here has been solely to quantify the effect of these.  A complete calculation including form factors and a longer list of reactions is a topic for a separate study.

Since Adler's theorem is relevant at low-energy, the
near-threshold cross sections are expected to be affected
the most.   We found the cross section for $J/\psi+\pi\rightarrow
(D^{*}+\bar{D}) + h.c.\,$ to be reduced as compared to   
a choice of Lagrangian which does not encode the full flavour 
chiral symmetry and does not obey Adler's theorem.  The reduction
is energy dependent, but seems to be a few tens of percents
from threshold to $\sqrt{s}$ = 5 GeV. In a full calculation the size of this reduction 
might not persist when one takes into account not only form factors, but also 
abnormal parity interactions and symmetry breaking effects (e.g. pseudoscalar masses 
and non-degenerate vector mass spectrum). 

Abnormal parity interactions may play an important role near threshold. Indeed, it was shown that Adler's theorem breaks down if a soft Goldstone boson can be inserted on an external line. This is expected to happen for abnormal parity Lagrangians where a $VV\phi$ vertex exists. The abnormal parity contribution to the $J/\psi +\pi$ amplitude will then not vanish in the soft limit. The problem in including these interactions lies again in the lack of experimental data to fix the coupling strengths. 

Symmetry breaking effects are also expected to be important since the underlying $SU_L(4)\times SU_R(4)$ is broken. Work is currently being done to include the physical mass of the vector meson within this formalism, while insisting that Adler's theorem hold for pions in normal parity interactions.

It will also be important in the
future to take this formalism to completion by implementing
covariant hadronic form factors computed within the same
effective Lagrangian or perhaps other approaches.  Ultimately, the outlook
for this line of study is to estimate the dissociation cross
sections with all the light hadrons, with finite size effects
incorporated, and then to input the results into a
dynamical model for heavy ion reactions to finally address
the question of $J/\psi$ survivability in the hadronic
phase (primarily mesonic matter).
For then, one would have a more complete understanding of the $J/\psi$ yield 
and therefore know what it implies about QGP formation.


\begin{acknowledgments}
A.B. thanks S.Turbide for helpful discussions, and A.B. and C.G. thank E. S. Swanson 
for  a useful visit. 
This work was supported in part by the Natural Sciences and Engineering 
Research Council of Canada, in part by the Fonds Nature et Technologies
of Quebec, and in part by the National Science Foundation under
grant number PHY-0098760. 
\end{acknowledgments}

\appendix
\section{Decoupling of the pion in the $\boldmath{\mbox{$J/\psi + \pi 
\rightarrow D^* + \bar D$}}$ process}
\label{appA}
Here we look at the soft-momentum limit of the full amplitude for the absorption 
process $J/\psi + \pi$ under the degenerate-vector meson mass condition. Again, 
we set one of the pseudoscalars' 4-momentum to zero (i.e. here $p_1$, but the 
proof is identical for $p_2$). As shown in a Section III B, for on-shell vector 
particles, the first amplitude goes to zero. For the second amplitude we have 
\begin{equation}
i\mathcal{M}^2_{\mu\nu} (p_1 \rightarrow 0) =  -ig_{J/\psi D^*D^*}g_{\pi DD^*}
\frac{[(2q_2-q_1)_\nu g_{\mu\lambda} + (2q_1-q_2)_\mu g_{\lambda\nu} -
(q_1+q_2)_\lambda g_{\mu\nu}]p_{2\lambda}(1+\Delta)}{(q_1-q_2)^2-m_D^{*2}}. \\
\end{equation}
Using momentum conservation in the first and second terms and noting that 
$t = (q_1-q_2)^2 = (p_2-p_1)^2$ yields
\begin{equation}
i\mathcal{M}^2_{\mu\nu} (p_1 \rightarrow 0) =  -ig_{J/\psi D^*D^*}
g_{\pi DD^*}\frac{[(q_2-p_2)_\nu g_{\mu\lambda} + (q_1+p_2)_\mu 
g_{\lambda\nu} -(q_1+q_2)_\lambda g_{\mu\nu}]p_{2\lambda}(1+\Delta)}
{p_1^2-2p_1\cdot p_2 + p_2^2-m_D^{*2}}, \\
\end{equation}
which reduces  to 
\begin{eqnarray}
i\mathcal{M}^2_{\mu\nu} (p_1 \rightarrow 0) &=&  \frac{ig_{J/\psi D^*D^*}
g_{\pi DD^*}}{m_{D^{*2}}-m_D^2}[q_{2\nu}p_{2\mu}-p_{2\nu}p_{2\mu} + 
q_{1\mu}p_{2\nu} + p_{2\mu}p_{2\nu} - (q_1+q_2)\cdot p_2 g_{\mu\nu} ]
(1+\Delta) \nonumber \\
&=& \frac{ig_{J/\psi D^*D^*}g_{\pi DD^*}}{m_{D^{*2}}-m_D^2}[q_{2\nu}p_{2\mu} 
+ q_{1\mu}p_{2\nu} - (q_1^2-q_2^2)g_{\mu\nu}  ](1+\Delta)\,.
\end{eqnarray}  
Contracting with the polarization vectors gives
\begin{eqnarray}
i\mathcal{M}^2(p_1 \rightarrow 0) &=& \frac{ig_{J/\psi D^*D^*}g_{\pi DD^*}}
{m_{D^{*2}}-m_D^2}(1+\Delta) \epsilon^{*\nu}(q_2)[ q_{2\nu}p_{2\mu} + 
q_{1\mu}p_{2\nu} - (q_1^2-q_2^2)g_{\mu\nu}] \epsilon^{\mu}(q1) \nonumber \\
&=&\frac{ig_{J/\psi D^*D^*}g_{\pi DD^*}}{m_{D^{*2}}-m_D^2}(1+\Delta) 
\epsilon^{*}(q_2)\cdot \epsilon(q_1) [m_{D^*}^2-m_{J/\psi}^2] \nonumber \\
&=& 0.
\end{eqnarray}
In the $SU_L(4)\times SU_R(4)$ model, the pseudoscalar meson thus decouples. But for 
the $SU_V(4)$, where $\mathcal{M}^3$ (contact term) is present, the full amplitude 
does not disappear. Note also that, as pointed out in \cite{nav01}, 
unless the underlying vector meson masses are degenerate (i.e. the vector mesons are 
arranged in $SU(4)$ multiplets) we have a residual contact term due to the second 
amplitude, and consequently the amplitude will not vanish when one of the pseudoscalar 
momenta goes to zero. 

\section{Electromagnetic current conservation for the $\boldmath{\mbox{$\gamma
+ \pi^+ \rightarrow \bar D^0 + D^{*+}$}}$ process}
\label{appB}

The proof that the electromagnetic current is conserved for this process in the 
$SU_V(4)$ model is given in \cite{hag00}. The authors invoke VMD, which we have 
shown to be exact in this model. For the $SU_L(4) \times SU_R(4)$ model, we have 
five amplitudes to consider (Fig.~\ref{fig4}): two which involve three intermediate 
particles (i.e. $\rho^0$, $\omega$, and $J/\psi$), two $s\,$-channel
contributions (one dominated by the $\rho$-meson and one through a 
direct $\gamma\pi\pi$ vertex), and a 4-point interaction.
\begin{figure}[!h]
\begin{center}
\includegraphics[scale=0.75]{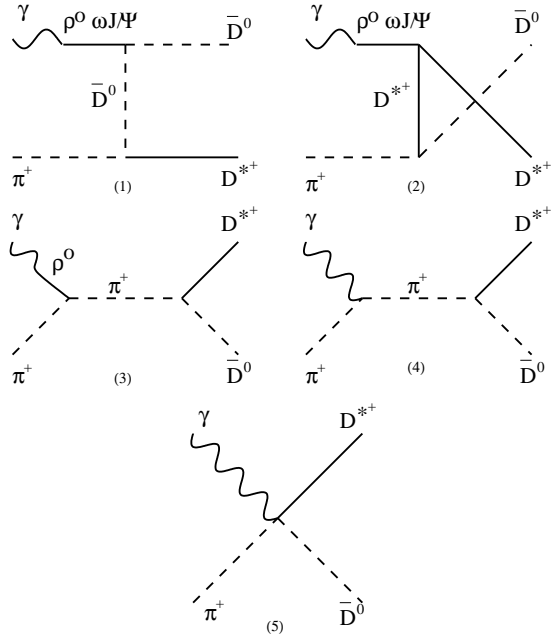}
\caption{Diagrams to be considered in the $SU_L(4) \times SU_R(4)$ model for the 
process $\gamma+ \pi^+ \rightarrow D^{*+} + \bar D^0$.}
\label{fig4}
\end{center}
\end{figure}
More specifically, the five amplitudes are
\begin{eqnarray}
\mathcal{M}_{\mu\nu}^{1} &=& -\frac{iekgF_\pi^{\,2}\,g^2_{V\phi\phi}}{2}\left[\frac{1}{4} 
+ \frac{1}{12}-\frac{1}{3}\right] \frac{[g_{\nu\alpha}-q_{1\nu}q_{1\alpha}/m_V^2]}{q_1^2-m_V^2} \frac{(2p_{2\alpha}-q_{1\alpha})(2p_{1\mu}-q_{2\mu})}{(q_1-p_2)^2-m_{D}^2} =0  
\nonumber \\
\mathcal{M}_{\mu\nu}^2 &=& -\frac{iekgF_\pi^{\,2}g\,g_{V\phi\phi}}{2}\left[\frac{1}{2}-
\frac{1}{6} + \frac{2}{3}\right] \frac{[g_{\nu\alpha}-q_{1\nu}q_{1\alpha}/m_V^2]}
{q_1^2-m_V^2} \nonumber \\ &&\left[(2q_2-q_1)_\alpha g_{\mu\lambda} + (2q_1-q_2)_\mu 
g_{\lambda\alpha} -(q_1+q_2)_\lambda g_{\mu\alpha}\right]\frac{p_{1\lambda}(1+\Delta)+
p_{2\lambda}(1-\Delta)}{(q_1-q_2)^2-m_{D^*}^2} \nonumber \\
\mathcal{M}_{\mu\nu}^{3} &=&  -\frac{iekgF_\pi^{\,2}\,g^2_{V\phi\phi}}{4} 
\frac{[g_{\nu\alpha}-q_{1\nu}q_{1\alpha}/m_V^2]}{q_1^2-m_V^2} \frac{(2p_{1\alpha}+q_{1\alpha})(2p_{2\mu}+q_{2\mu})}{(q_1+p_1)^2-m_\pi^2} \nonumber \\
\mathcal{M}_{\mu\nu}^4 &=& \frac{ie(1-k/2)g_{V\phi\phi}}{2}\frac{(2p_{1\nu}+q_{1\nu})(2p_{2\mu}+q_{2\mu})}{(p_1+q_1)^2-m_\pi^2} \nonumber  \\
\mathcal{M}_{\mu\nu}^5 &=& -\frac{ieg_{V\phi\phi}}{2}g_{\mu\nu}, \nonumber
\end{eqnarray}
where the first amplitude vanishes because of the $SU(4)$ structure of the vector 
meson multiplet. Contracting with $q_{1\nu}$ and $\epsilon^*_\mu(q_2)$ we find
\begin{eqnarray}
\epsilon^{*\mu}(q_2)q^\nu_{1}\mathcal{M}_{\mu\nu}^{1} &=&-ie\frac{kg_{V\phi\phi}}
{2}\left[\frac{1}{4}+\frac{1}{12}-\frac{1}{3}\right]2p_1 \cdot \epsilon^*(q_2)   = 
0  \nonumber \\
\epsilon^{*\mu}(q_2)q^\nu_{1}\mathcal{M}_{\mu\nu}^2 &=& \frac{ie2g_{V\phi\phi}}{2}
\left[-\frac{1}{4}+\frac{1}{12}-\frac{1}{3}\right] (p_1+p_2)\cdot \epsilon^*(q_2) =
-\frac{ieg_{V\phi\phi}}{2} (p_1+p_2)\cdot \epsilon^*(q_2) \nonumber \\
\epsilon^{*\mu}(q_2)q^\nu_{1}\mathcal{M}_{\mu\nu}^{3} &=& +\frac{iekg_{V\phi\phi}}
{4}2p_2\cdot \epsilon^*(q_2) \nonumber \\
\epsilon^{*\mu}(q_2)q^\nu_{1}\mathcal{M}_{\mu\nu}^4 &=& \frac{ie(1-k/2)
g_{V\phi\phi}}{2} 2p_2\cdot \epsilon^*(q_2) \\ \nonumber  
\epsilon^{*\mu}(q_2)q^\nu_{1}\mathcal{M}_{\mu\nu}^{5} &=& -
\frac{ie2g_{V\phi\phi}}{2}\left[\frac{3}{4} + \frac{1}{12} -\frac{1}{3}\right] 
q_{1}\cdot \epsilon^*(q_2) = -\frac{ieg_{V\phi\phi}}{2}q_{1}\cdot \epsilon^*(q_2), 
\nonumber 
\end{eqnarray}
where $kg = g_{V\phi\phi}$ and $gF_\pi^{\,2}/m_V^2 = 1/g_{V\phi\phi}$. Using momentum 
conservation we see that the Ward identity holds when we add up all the contracted 
amplitudes. In the case where $k=2$ the Ward identity can be shown to hold for each 
subset of diagrams with a particular intermediate vector particle as in \cite{hag00}.

\section{Fixing coupling constants}
\label{appC}

Since the purpose here is only to compare cross sections calculated 
within two models, form factors will not be introduced. Clearly, in a 
complete calculation, these would have to be included. Furthermore, the 
coupling constants will be fixed by fitting $\rho$ phenomenology and
then using symmetry relations. 
For the interaction term 
\begin{equation}
\mathcal{L} = -i\frac{g_{\rho\pi\pi}}{2}{\rm Tr}\left[\rho^\mu[\pi,\partial_\mu 
\pi]\right],
\end{equation}
the corresponding width is
\begin{equation}
\Gamma(\rho \rightarrow \pi\pi) = \frac{g_{\rho\pi\pi}^2}{12\pi} \frac{|p_\pi|^3}
{m_\rho^2}.  
\end{equation}
With the measured width of $\Gamma(\rho \rightarrow \pi\pi) = 151$ MeV and $\rho$ 
and $\pi$ masses of $m_\rho = 770$ MeV and $ m_\pi = 140$ MeV, the coupling constant 
is evaluated at $g_{\rho\pi\pi} = 8.55$. Noting that $g_{SU_V(4)} = g_{\rho\pi\pi}$ 
and $g_{SU_L(4)\times SU_R(4)} = m_\rho^2/g_{\rho\pi\pi}F_\pi^{\,2} = 3.98$ ($F_\pi = 132$ 
MeV), and using the symmetry relations, all the coupling constants can be evaluated 
(see Table I). Besides the difference in the contact term,  the slight 
difference between the two models for the coupling $g_{J/\psi D^*D^*}$ is
attributable to the presence of the extra parameter $k$ ($ = 2.15$) in the 
$SU_L(4)\times SU_R(4)$ Lagrangian. 
An alternate approach is to fix the coupling constants by fitting known hadronic and 
radiative decay widths using VMD \cite{mat95, lin00,oh01}. The symmetry is 
then invoked for determining the 4-point coupling for which there is no specific empirical 
information.
\begin{table}[!h] 
\begin{tabular}{|c|c|c|}
\hline
Coupling constant & $SU_V(4)$  & $SU_R(4) \times SU_L(4)$ \\
\hline
$g_{J/\psi DD}$ & $4.94$  & $4.94$\\
$g_{J/\psi D^*D^*}$ &$4.94$& $4.60$ \\
$g_{\pi DD^*}$ &$4.28$&$4.28$ \\
$g_{J/\psi D^* D\pi}$ &$21.10$ & 0 \\
\hline
\end{tabular}
\caption{Coupling constants for the the two models considered for the $J/\psi +\pi$ 
absorption process.}
\label{tab1}
\end{table}


\begin{thebibliography}{99}
\bibitem{kl03} F. Karsch and E. Laermann, {\it Quark-Gluon Plasma 3}, edited by 
Rudolph C. Hwa and 
Xin-Nian Wang (World Scientific, Singapore, 2004).
\bibitem{mat86} T. Matsui and H. Satz, Phys. Lett.{\bf B178}, 416 (1986).
\bibitem{ks94} D. Kharzeev and H. Satz, Phys. Lett. {\bf B334}, 155 (1994).
\bibitem{na38} M. C. Abreu {\it et al.}, Phys. Lett. {\bf B449}, 128 (1999).
\bibitem{na50} M. C. Abreu {\it et al.}, Phys. Lett. {\bf B477}, 28 (2000).
\bibitem{arm98} N. Armesto and A. Capella, Phys. Lett. B {\bf 430}, 23 (1998).
\bibitem{cap02} A. Capella, A. B. Kaidalov, and D. Sousa, Phys. Rev. C 
{\bf 65}, 054908 (2002).
\bibitem{dur03} D. Kharzeev, H. Satz, A. Syamtomov, and G. Zinovjev, Phys.
Lett. {\bf B389}, 595 (1996);
F.~O.~Duraes, H.~C.~Kim, S.~H.~Lee, F.~S.~Navarra, and 
M.~Nielsen, Phys.\ Rev.\ C {\bf 68}, 035208 (2003).
\bibitem{mar95} K. Martins, D. Blaschke, and E. Quack, Phys. Rev. {\bf 51}, 
2723 (1995).
\bibitem{won01} C.-Y. Wong, E.S. Swanson, and T. Barnes, Phys. Rev. C 
{\bf 65}, 014903 (2001); 
\bibitem{mat95} S. G. Matinyan and B. M\"uller, Phys. Rev. C {\bf 51}, 2723 
(1995).
\bibitem{kh00} K. L. Haglin, Phys Rev C {\bf 61}, 031902R (2000).
\bibitem{hag00} K. L. Haglin and C. Gale, Phys Rev C {\bf 63}, 065201 (2001).
\bibitem{lin00} Z. Lin, C.M. Ko, and B. Zhang, Phys. Rev. C {\bf 61}, 024904 
(2000).
\bibitem{oh01} Y. Oh, T. Song, and S. Houng Lee, Phys. Rev. C {\bf 63}, 
034901 (2001).
\bibitem{nav01} F. S. Navarra, M. Nielson, and M. R. Robilotta, Phys. Rev C 
{\bf 64}, 021901 (2001).
\bibitem{mei88} U. Meissner, Phys. Rept. {\bf 161}, 215 (1988). 
\bibitem{kay85} \"O. Kaymakcalan and J. Schechter, Phys. Rev. D {\bf 31}, 1109
(1985).
\bibitem{gom84} H. Gomm, \"O. Kaymakcalan, and J. Schechter, Phys. Rev. D 
{\bf 30}, 2345 (1985).
\bibitem{gai69} S. Gasiorowicz and  D. A. Geffen, Rev. Mod. Phys. {\bf 41}, 531 
(1969).
\bibitem{wei68} S. Weinberg, Phys. Rev. {\bf 166}, 1568 (1969).
\bibitem{sak69} J.J. Sakurai, {\it Currents and Mesons}, University of 
Chicago Press, Chicago, (1969).
\bibitem{sch86} J. Schechter, Phys. Rev. D {\bf 34}, 868 (1986).
\bibitem{wei95} S. Weinberg, {\it The Quantum Theory of Fields}, Vol. 2,
Cambridge University Press, (1995). 
\bibitem{adl65} S. Adler, Phys. Rev. {\bf 137}, B1022 (1965).
\bibitem{bur00} C.P. Burgess, Phys. Rept. {\bf 330}, 193-261 (2000).
\bibitem{wit83} E. Witten, Nucl. Phys. B {\bf 223}, 422 (1983).
\bibitem{kro67} N. M. Kroll, T.D. Lee, and B. Zumino, Phys. Rev. {\bf 157}, 
1376 (1967).
\bibitem{not01} The apparent discrepancy between Eq. (3a) of \cite{oh01} and 
Eq. (7) of \cite{lin00} is due to a typographical error, as the Lagrangian of the former
(Eq. (A5)) and that of the latter (Eq. (6)) are  formally identical.
\bibitem{lin002} Z. Lin, C. M. Ko, and B. Zhang, Phys. Rev. C {\bf 61}, 024904 (2000). They point out that their model is motivated by the hidden gauge approach where there is no four-point vertex. They argue that inclusion of the charm axial vectors, which are unknown, would make the models agree.

\bibitem{cleo} A. Anastassov {\it et al.}, Phys. Rev. D {\bf 65}, 032003 (2002)

\end{thebibliography}
\end{document}